\documentclass[reprint,aps,prx,amsmath,amssymb,twocolumn,nobalancelastpage,citeautoscript,floatfix,superscriptaddress,longbibliography]{revtex4-2}
\usepackage[usenames,dvipsnames]{color}
\usepackage{graphicx,xfrac,microtype}
\usepackage[bookmarks=false,colorlinks]{hyperref}
\hypersetup{hypertexnames=false,linkcolor=RubineRed,citecolor=NavyBlue,filecolor=Mulberry,urlcolor=RoyalBlue}
\newcommand{\Autoref}[1]{
  \begingroup%
  \def\sectionautorefname{Appendix}%
  \autoref{#1}%
  \endgroup%
}
\usepackage{enumitem} 
\setlist{leftmargin=*}

\newcommand{\revision}[1]{\textcolor{black}{#1}}

\begin{document}

\title{Learning the Crystal Structure Genome for Property Classification}

\author{Yiqun Wang}
\affiliation{Department of Materials Science and Engineering, Northwestern University, Evanston, Illinois 60208, USA}

\author{Xiao-Jie Zhang}
\affiliation{Department of Chemistry, Fudan University, Shanghai, China}
  
\author{Fei Xia}
\affiliation{Department of Electrical Engineering, Stanford University, Stanford, California 94305, USA}

\author{Elsa A.\ Olivetti}
\affiliation{Department of Materials Science and Engineering, Massachusetts Institute of Technology, Cambridge, Massachusetts 02139, USA}

\author{Stephen D.\ Wilson}	
\affiliation{Materials Department, University of California, Santa Barbara, Santa Barbara, California 93106, USA}

\author{Ram Seshadri}
\affiliation{Materials Department, University of California, Santa Barbara, Santa Barbara, California 93106, USA}

\author{James M.\ Rondinelli}
\email{jrondinelli@northwestern.edu}
\affiliation{Department of Materials Science and Engineering, Northwestern University, Evanston, Illinois  60208, USA}

\begin{abstract}
%
\noindent Materials property predictions have improved from advances in machine learning algorithms, 
delivering materials discoveries and novel insights through data-driven models of structure-property relationships. 
Nearly all available models rely on featurization of materials composition, 
however, whether the exclusive use of structural knowledge in such models 
has the capacity to make comparable predictions remains unknown. 
Here we employ
\revision{a deep neural network model}
to 
\revision{decode}
structure-property relationships in crystalline materials 
without explicitly considering chemical compositions. 
The focus is on classification of crystal systems, mechanical elasticity, 
electronic band gap, and phase stability. 
Our model utilizes a three-dimensional (3D) momentum space representation of structure from elastic X-ray scattering theory 
that exhibits rotation and permutation invariance.
\revision{We perform novel ablation studies to help interpret the model performance by perturbing the physically meaningful input features (i.e., the diffraction patterns) instead of tuning the architecture of the learning model as in conventional ablation methods.
We find that the spatial symmetry of the 3D diffraction patterns, which reflects crystalline symmetry operations, is more important than the 
diffraction intensities contained within for 
the model to make a successful classification.
Our work showcases the potential of using statistical learning models to help understand materials physics, 
rather than performing predictive and generative tasks as in most materials informatics research.}
%
%
We also argue that
learning the crystal structure genome in 
a chemistry-agnostic manner 
demonstrates that some crystal structures inherently host high propensities for optimal materials properties, 
which enables the decoupling of structure and composition for future co-design of multifunctionality. \\[-0.75em]

\end{abstract}
\maketitle

\begin{figure*}
  \centering
  \includegraphics[width=0.98\linewidth]{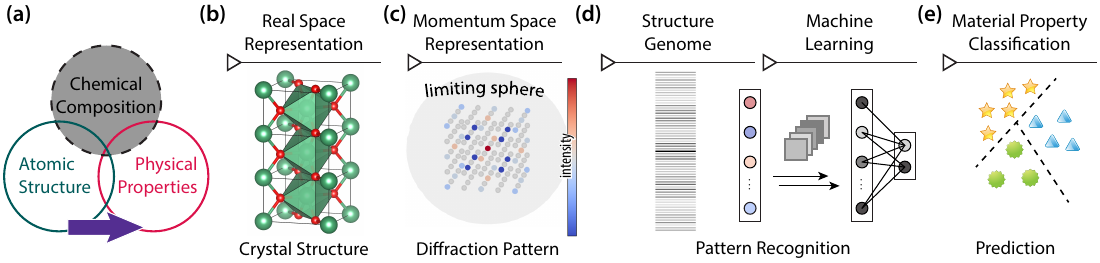}
  \caption{%
  Workflow for constructing
  a machine learning model to learn (a) structure-property relations without featurization of chemical composition. The process begins with (b) the real-space crystal structure representation (in either the conventional cell or primitive cell), which is transformed into a (c) momentum space representation by simulating the 3D X-ray diffraction patterns, which is represented as point cloud. Only diffraction points within the limiting sphere are physically observable. (d) We then construct the model to learn directly from the point cloud data to accomplish (e) property-classification tasks.
  }
  \label{fig:fig1}
\end{figure*}

\section{Introduction}

One of the most frequently used phrases in materials research is 
``structure-property relationships.'' 
It forms the cornerstone of forward and inverse system-level-based materials design \cite{Olson2000,Xiong2016}, 
and it is principally used in two modalities: (1) to exclusively describe relationships 
for a single material family, such that the composition is fixed, 
and dependencies arise from processing-based microstructural changes, 
or (2) to explicitly describe effects arising from  changes in composition, 
which  inadvertently contracts the full ``structure-composition-property" relationship phrase 
despite chemical dependencies dominating structural changes.  
Admittedly, both atomic structure and chemistry 
mutually determine materials properties (\autoref{fig:fig1}a).
The intimate interwoven description of what defines a material -- 
the elemental species involved and the crystallographic structure the atoms adopt once bonded together 
given a fixed ratio -- and which 
physical properties can ``live'' in various structures  
pose a challenge for novel materials design and discovery.  
With the absence of theoretical or statistical guidance, 
materials scientists need to search 
through a combinatorial
space spanned by both chemical compositions as well as structure types \cite{Rondinelli2015}.

Despite the key role chemistry plays in physical properties, condensed-matter physicists have 
harnessed effective theoretical models, 
e.g., Hubbard, Heisenberg, and Fu-Kane models, etc., 
based on different interactions, orbital symmetries, 
and topologies to describe the electronic and magnetic phases of 
materials without explicitly encoding material composition.
The premise relies on recognition that the low-energy 
electrons comprising atoms interact on a lattice, which may be mapped 
onto a (portion of a) known crystal structure.
Even with modern computational simulations, e.g., those 
based on density functional theory (DFT), chemical information is 
only included in the form of atomic orbitals at each crystallographic site and their corresponding atomic numbers to provide a potential for the 
electrons to interact.
To that end, we pose the following question:
\textsl{Is it possible to marginalize compositional information and understand to what extent crystal structure exclusively determines materials properties?}

In this work, we address this question using a statistical learning-based method, 
leveraging open access to 
numerous materials databases  \cite{Jain2013,Saal2013MaterialsOQMD,Curtarolo2012,Borysov2017} 
and recent advances in materials informatics tools \cite{butler2018machine,ramprasad2017machine,agrawal2016perspective,chen2021learning}.
Many machine learning (ML) models exploiting these data  
have successfully
predicted materials properties:  
local connectivity-based models \cite{isayev2017universal} 
and graph neural networks \cite{xie2018crystal,chen2019graph,karamad2020orbital} 
have achieved DFT-level performance,
and have helped accelerate the discovery of novel functional materials \cite{ahmad2018machine,ren2020inverse,Georgescu2021}.
Here, we learn the materials structure-property relationship 
from crystal structure alone -- without use of chemical composition 
as illustrated in \autoref{fig:fig1}a 
-- to 
\revision{make classifications on} 
a variety of properties including crystal system, 
elasticity, metallicity, and stability.
%
%

\revision{Most existing materials informatics models typically utilize both structural and compositional information as features in order to achieve high predictive performance.}
\revision{In contrast, the goal of this study is to understand the relative contribution and significance of crystal structures to materials properties, which is enabled by learning from ``incomplete'' materials structural descriptors without chemical compositions.
Specifically, }
we use a momentum-space representation of crystal structures in 
the form of simulated X-ray diffraction (XRD) patterns to generate a three-dimensional (3D) point cloud, 
which serves as a unique structural fingerprint of each material.
We then construct and train a deep neural network (DNN), which is invariant
under rotation and permutation operations on the input 3D XRD patterns, 
\revision{to perform property classifications.}
By concealing and perturbing information in the 3D point cloud fed to the DNN, 
we ascertain that crystal structure plays a decisive role in 
\revision{all aforementioned property prediction tasks, yet the relative significance is unique to each material property.
We also achieve similar predictive performance using other machine learning classification models using the same descriptors, 
from which we argue that our findings are general, 
so that one should focus more on the resulting analyses (i.e., physical interpretations) rather than the model architecture (i.e., opening the black box).}
Our findings reveal the correlation between crystal structures and multiple common materials properties, 
which could enable co-design of 
\revision{functional materials}
by prioritizing optimization of crystal structure or composition to achieve 
desired performances.

\begin{table*}
\caption{\label{tab:table1}%
The \revision{machine learning model} is trained on the 3D point cloud representation of materials from the Materials Project database, but the number of materials used for different classification tasks varies due to data availability. The classification boundary values are chosen so as to ensure class balance. \revision{$E_g$, $B$, $H$, and $E_H$ represent the  electronic band gap, bulk modulus, shear modulus, and energy above the convex hull, respectively.}}
\begin{ruledtabular}
\begin{tabular}{lll}
    Classification Task & Total Compounds  & Class Distribution  \\
\hline
Metallicity & 74,108 & 36,451  with $E_g = 0$\,eV (metal) and 37,657 with $E_g > 0$\,eV (insulator) \\
Elasticity  & 13,176 & 6,582 with $B>85$\,GPa and 6,472 with $G>34$\,GPa \\
Phase Stability & 139,029 & 58,027 stable compounds with $E_H<20$\,meV\,atom$^{-1}$ \\
\end{tabular}
\end{ruledtabular}
\end{table*}

\section{Methodology}

\subsection{Materials Representation}\label{sec:materials_representation}

A perfect crystal under periodic boundary conditions in real space is 
mathematically described as the convolution of its Bravais lattice (BL) and the 
atomic structure of the asymmetric unit (motif) within the unit cell 
(\autoref{fig:fig1}b).
Owing to the periodicity in real space, materials scientists typically use diffraction-based methods (e.g., X-ray or neutron scattering) to determine the crystal structures. 
The process of X-ray diffraction is the mathematical equivalence of a Fourier transform 
($\mathcal{F}$); 
it converts the real-space crystal structure into momentum space and forms a new reciprocal-space lattice  exhibiting intensities dependent on the so-called structure factor ($F$) as: 
\begin{eqnarray}
	 \mathcal{F}(\textrm{BL}* \textrm{motif}) &=&  \mathcal{F}(\textrm{BL})\cdot \mathcal{F}(\textrm{motif}) \nonumber \\
	&=& (\textrm{reciprocal lattice}) \cdot F_{hkl}
	\label{eq:eq1}
\end{eqnarray}
where $*$ and $\cdot$ are the convolution and product operations, respectively, and $h\,k\,l$ are integer labels of the reciprocal lattice points that correspond to the Miller indices for lattice planes in real space. 
The aforementioned real-space convolution relationship then becomes 
a product between the reciprocal lattice and structure factor $F_{hkl}$
\revision{based on the convolution theorem}. 
The physical observable from XRD is the diffraction intensities $I_{hkl}$ (real), not the structure factors $F_{hkl}$ (complex).
Rather, $I_{hkl}$ is proportional to the square modulus of the structure factor 
$|F|^2=F^{*}_{hkl} \cdot F_{hkl}$, where $*$ is the  complex conjugate.
$F_{hkl}$ serves as the Fourier series coefficients of the real space periodic electron density $\rho(\mathbf{r})$ derived from atoms located at $\mathbf{r}_j$ in the unit cell: 
\begin{equation}
F_{hkl} = \frac{1}{V_\mathrm{cell}}\sum_{j=1}^{N} f_j(\mathbf{g}_{hkl}) e^{2\pi i (\mathbf{g}_{hkl} \cdot \mathbf{r}_j)}\,, \label{eq:eq2} 
\end{equation}
where $f_j(\mathbf{g}_{hkl})$ represents the atomic scattering factor for atom $j$ at reciprocal point $\mathbf{g}_{hkl}$:
\begin{equation}
	f_j(\mathbf{g}_{hkl}) = \int \!d\mathbf{r}_j \,\rho(\mathbf{r}_j) e^{2\pi i(\mathbf{g}_{hkl}\cdot \mathbf{r}_j)} \,. \label{eq:eq3}
\end{equation}

Given the intensity $I_{hkl}$ encodes both atomic structure and electron density information, we propose to utilize it as a
3D momentum space representation
for predicting physical properties of crystalline materials 
without explicit compositional features.
The diffraction intensity values reflect the number of electrons associated with an ion or element in a material.
Owing to the \revision{infamous} \textit{phase problem} in crystallography -- the complex phase factor is lost upon calculating the square modulus of $F_{hkl}$ --
reconstructing the original electron density function through a direct inverse Fourier transform, however, is not feasible.
Chemical composition \revision{identification/inference} is then nearly impossible for our model.
The spatial distribution of diffraction intensities, however, 
are unique to each material as they depend on crystal symmetries of the atomic structure 
\footnote{It is possible to artificially make two materials exhibit identical diffraction patterns, but we only consider materials in equilibrium states}. 
Therefore, we use the intensity distribution as the structural signature 
from which to \revision{make classifications on} materials properties.
Since the mapping functions from the diffraction intensity $I_{hkl}$
to the target materials properties are unknown (\autoref{fig:fig1}a, purple arrow), 
we use \revision{machine learning models} to decode the structure-property relationship as they are ideal candidates for function approximation \revision{given enough training data}.
Owing to the fact that existing experimental methods typically access a 2D slice of the full 3D diffraction patterns, 
and not all experimental XRD patterns are readily available in open databases,
we simulate the full 3D patterns using a modified version of the XRD calculator implemented in \texttt{Pymatgen} \cite{ong2013python}
\revision{, the implementation details are available at the project \href{https://github.com/raymond931118/deepKNet}{GitHub} page.}

We retrieved materials data from the Materials Project database \cite{*[{}][{ using data retrieved on October 30, 2021.}] Jain2013}.
%
%
%
We obtained a dataset comprising 
\revision{139,367} materials with the following specified properties: 
crystal system, bulk modulus ($B$), shear modulus ($G$), electronic band gap ($E_g$), and energy above the convex hull ($E_H$).
All materials properties utilized herein 
were simulated using DFT by the Materials Project.
Since not all properties are available for every compound in the database,
the total number of materials for each classification task differs (\autoref{tab:table1}).
We assigned thresholds as shown in \autoref{tab:table1} for different classification tasks to ensure 
physically meaningful class boundaries (e.g., metals and insulators), 
while maintaining a well-balanced dataset.
\revision{For instance, the median value of bulk modulus (85\,GPa) was used to partition the dataset into two classes with different stiffness for binary classification.}

For each material, we first construct its 
\revision{primitive} standard cell
using the DFT-relaxed crystal structure reported by the Materials Project. 
\revision{(A conventional standard cell can also be used to achieve comparable performance, 
however, we report property classification results here using the primitive cell. It is the physical structural unit for the scattering process. In addition, using experimental crystal structures should lead to the same statistical results as reported later.)
}
Then, we simulate its 3D XRD pattern using 
Cu\,K$_\alpha$ radiation ($\lambda=1.5418$\,\AA).
Under our kinematic approximation, only reciprocal lattice points ($h\,k\,l$)
within the limiting sphere of radius $4\pi /\lambda$ 
exhibit finite diffraction intensity 
while the intensity in the remainder of momentum space is strictly  zero (\autoref{fig:fig1}c).
The initial features for each material then comprise a set of 
$\{[h_i, k_i, l_i, I_i] \mid  i \in [1, n]\}$ diffraction points, 
where $n$ is the total number of points 
within the limiting sphere.
Since the shape and size of the reciprocal lattice 
vary from material to material, as they are dependent on the crystalline symmetry 
and real space lattice constants, 
each compound exhibits \revision{unique}
(1) diffraction point ($\mathbf{g}_{hkl}$) density, 
(2) configuration of these points within the limiting sphere, and 
(3) intensity values of these points.
\revision{In other words, using ($h\,k\,l\,$) indices alone may not be a good descriptor for structures across different crystal systems; the ordering of basis vectors can be defined arbitrarily. (One could permute randomly the three reciprocal basis vectors, which leads to a different ($h\,k\,l\,$) indexing, yet the underlying crystal remains invariant.)}
Therefore, we further convert the ($h\,k\,l\,$) indices of each diffraction pattern 
to Cartesian coordinates using the reciprocal lattice vectors.
\revision{The locations of diffraction points are thus described explicitly within the reciprocal space.}
We also take the natural log of the intensity values, $\ln(I+10^{-6})$, and normalize them 
to bring all features to a similar scale (i.e., within [0, 1]).
Implementation details are available in \Autoref{sec:derivations}.

\begin{figure}[t]
  \centering
  \includegraphics[width=0.9\columnwidth]{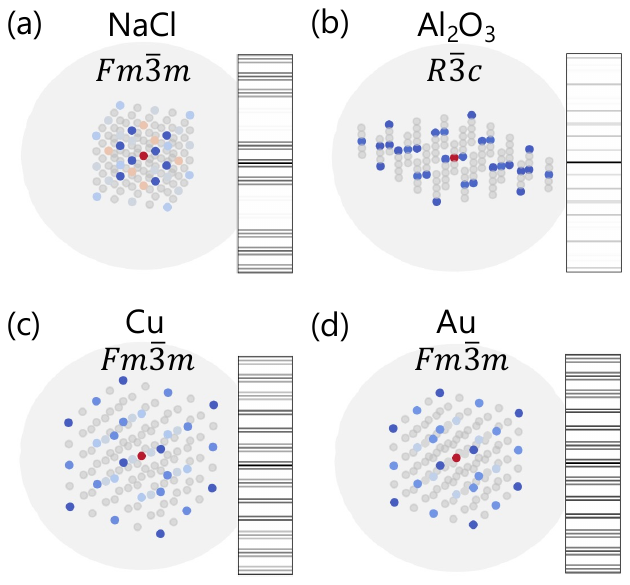}
  \caption{%
The simulated X-ray diffraction patterns of select crystals with corresponding space group. The gray spheres represent the limiting sphere of radius $\frac{4\pi}{\lambda}$. Only diffraction points with Miller indices within \{$\bar{2}$, $\bar{1}$, 0, 1, 2\} are shown here. The intensity of the origin (0\,0\,0) is calculated as the \revision{squared} total electron density within the cell. This point cloud representation of crystal structures simultaneously displays rotation and permutation invariance.
The centrosymmetric ``bar code" patterns are generated by sorting the diffraction patterns with respect to their corresponding Miller indices.
}
\label{fig:fig2}
\end{figure}

Since each material has a different diffraction point density within the limiting sphere, 
we define a fixed number of $\mathbf{g}_{hkl}$ points $n$ to featurize all compounds.
We discuss the impact of $n$ on model performance later.
Note that $n$ is a variable from which we can 
\revision{infer} materials physics; 
it is not a hyperparameter for 
\revision{tuning} 
the machine learning model.
\revision{Naively, a larger $n$ contains more information (i.e., lattice planes with smaller distance, higher structural resolution),
and we are interested in inspecting the model performance as a function of the amount of information presented to it.}
To that end, we specifically consider five different $n$ values, which is determined by the range of Miller indices included in the feature set:
\begin{itemize}
    \item Reciprocal points (1\,0\,0),\,(0\,1\,0),\,(0\,0\,1), $n=3$ points;
    \item Reciprocal points (0\,0\,0),\,(1\,0\,0),\,(0\,1\,0),\,(0\,0\,1), $n=4$ points;
    \item Miller indices $h\,k\,l\in\{\bar{1}, 0, 1\}$, $n=27$ points;
    \item Miller indices  $h\,k\,l\in\{\bar{2}, \bar{1}, 0, 1, 2\}$, $n=125$ points; and
    \item Miller indices $h\,k\,l\in\{\bar{3}, \bar{2}, \bar{1}, 0, 1, 2, 3\}$, $n=343$ points.
\end{itemize}
For instance, in the $n=125$ case, we include all combinations of $h\,k\,l$ within $\{\bar{2}, \bar{1}, 0, 1, 2\}$, for a total of 125 points, into the feature set.
The point cloud representation of some common crystals are shown in \autoref{fig:fig2}, from which we can see the diversity in point density, shape, and diffraction intensity across different materials (all exhibiting centrosymmetry, without considering anomalous scattering which breaks Friedel's law).
\revision{Note that here we show the diffraction patterns from  conventional standard cells to help visualize the features, while the model learns from diffraction patterns of the primitive cells. Our preliminary studies confirmed the choice of unit cell does not significantly impact the statistical results.}
All diffraction points beyond the considered index range are eliminated, and hence invisible to the model.
After this data pre-processing step,
all materials should have a feature set defined by an $n\times4$ array, with $n$ rows and 4 columns:
$[x, y, z, I]$, which represent the Cartesian coordinates and the log diffraction intensity, respectively.

This 3D crystalline material representation is in the form of point cloud---an unordered set of points distributed in high-dimensional space. 
Since the orientation of the reciprocal lattice basis is arbitrary, 
and the set of points do not follow a specific order, 
rotating the point coordinates 
or swapping the order of two points 
should not have any impact on material properties.
This behavior is different from pixels in an image.
Therefore, our model should be invariant under both 3D rotation 
and permutation operations on the input points.
In order to enforce the rotation and permutation invariance of our model, we apply random 3D rotation and random shuffling of the point sequence of each material before feeding them to the model.
These data-augmentation transformations are performed on-the-fly (i.e., in-memory operation after loading the original features) instead of being precomputed and stored on hard drive.
Specifically, we use 3 randomly and independently generated Euler angles within the range $[-\frac{1}{4}\pi, \frac{1}{4}\pi]$ for the crystal system classification task, while we use $[-\pi, \pi]$ for all physical property classification tasks. The justification for selecting different ranges of the Euler angles is explained later (\emph{vide infra}).
%
Therefore, the model never sees the same representation of a material twice, yielding an effectively 
infinitely sized dataset.
In addition, we show later that the performance of the model for property classifications on the test dataset is independent of the random 3D rotations and point permutations.

We split the dataset into training, validation, and test sets, 
with ratios of 70\%, 15\%, and 15\%, respectively.
The validation set is used to select the optimal model hyperparameters.
We report the model performance on the test set 
containing materials that the model has never seen.
Since our goal is to understand materials physics using a machine learning model as an information extractor 
\revision{(rather than achieving the best predictive performance)},
we train each model on 3 randomly and independently generated training-validation-test datasets, 
and report 
\revision{the mean values of performance metric 
(and their standard deviations)}
on the test set to reduce the variance of results.

\subsection{Model Architecture}

\begin{figure*}
  \centering
 \includegraphics[width=0.98\textwidth]{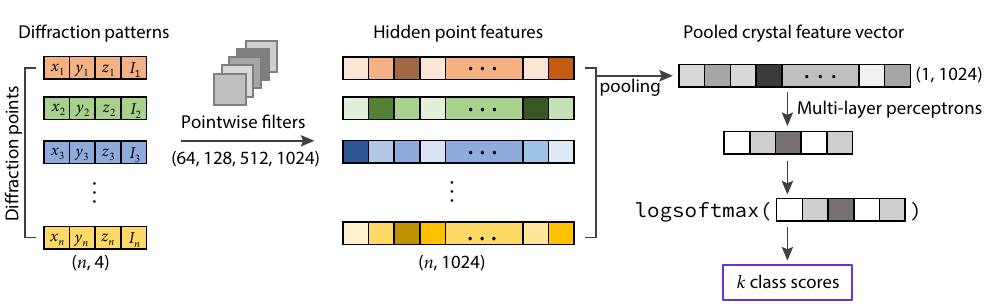}
  \caption{The neural network architecture. 
  Multiple pointwise filters (multilayer perceptrons)
  are applied to extract the position-intensity relationship from the simulated diffraction data.
  The shape of feature and intermediate tensors are indicated in parenthesis and $n$ is the number of diffraction points 
  \revision{based on the range of Miller indices considered}. 
  Operations in this step do not involve point-point communications; therefore, permutation invariance is preserved. Then, a symmetric function is used to pool the crystal feature vector from all diffraction points. Here, the \texttt{max} pooling function is used, but others also work.
  Lastly, multi-layer perceptrons are used to eventually make the classification decision. See \Autoref{sec:hyperparameter_optimization} 
  for details of model hyperparameter selection.}
  \label{fig:fig3}
\end{figure*}

Learning from 3D point-cloud data is an active area of computer-vision research. 
Owing to the rotation and permutation invariance requirements of our $I_{hkl}$ point-cloud representation, 
most conventional ML models cannot be directly applied to our 
learning problem.
For instance, conventional 2-dimensional convolutional neural networks (CNNs), 
which are the most prevalent network structure for 2D image classification tasks \cite{deng2009imagenet},
are robust against object translations; however, 
permutation of the input data (e.g., swapping pixels of an image) could break down the network.
Existing solutions to this problem include PointNet-based models \cite{qi2017pointnet,qi2017pointnet++,thomas2019kpconv}, multi-view CNN \cite{qi2016volumetric}, and some other CNN variants \cite{wu20153d,bruna2013spectral};
however, these tend to focus on object detection/classification and segmentation learning tasks.

Here we demand more from the neural network model, 
which goes beyond the 3D computer vision problem---the 
analogue of which would be identifying the 1 amongst 7 crystal 
systems a material belongs to by knowing how atoms are arranged in a unit cell.  
The features we use for the materials-property classification tasks include not only positional data (i.e., Cartesian coordinates), but also the diffraction intensity as the fourth dimension.
Thus, the input features together contain information about the 
cell shape, cell size,  symmetry, and electron density.
This information is all simultaneously
embedded within the sparse distribution of diffraction points in momentum space.
To that end, the DNN needs to learn 
the patterns of different material properties (e.g., metals and insulators) 
using their structural fingerprints, 
and not only to identify structural patterns given structural features  \cite{ziletti2018insightful}. 
%

The network architecture capable of solving this problem is 
elegant in its simplicity  
as depicted in \autoref{fig:fig3}.
Inspired by PointNet, we  
\revision{first extract pointwise features using 
multilayer perceptrons (MLPs), where
each feature column, i.e., Cartesian coordinates and intensity, 
is treated as one input channel, 
and the filters mix over features from different channels.
We apply batch normalization and dropout to the MLP layers as regularization.}
\revision{After pointwise feature extraction,}
the model learns the position-intensity relationship of different points, 
whose output features should be invariant to rotation of the Cartesian coordinates 
of input points (e.g., distance to origin).
The aforementioned steps only involve operations within each individual point, 
\revision{in other words,}
no point-point communications are made, 
hence preserving permutation invariance.
Now, the learned material representation becomes a tensor of shape ($n$, $m$), where $m$ is a hyperparameter indicating the number of embedding dimension. (We use $m=1024$ for all classification tasks.)

Then, we apply a symmetric function to aggregate information from all points.
We find that the \texttt{max} pooling function works well in all our tasks, and this operation safely preserves permutation invariance, because it does not involve point indexing.
\revision{Specifically, for each hidden feature column, the \texttt{max} pooling operation only takes the point with the highest value (i.e., the critical point) and neglects all other points which do not contribute to the pooled crystal feature vector (\autoref{fig:fig3}).}
In addition, we also tried a self-attention-based pooling algorithm, 
and found that the performance gain is negligibly small 
(e.g., ROC-AUC value from 0.910 to 0.915 for metal-insulator classification) 
while the model size became several times larger than using the \texttt{max} pooling function.
Therefore, although knowing that \texttt{max} pooling is not the only working method for information aggregation, we use this pooling function for all our classification tasks.
\revision{\texttt{Max} pooling} also enables physically meaningful model interpretation since it allows us to know which points contribute to the pooled crystal feature vector 
\revision{(by taking an \texttt{argmax} operation, as shown in \autoref{sec:model_interpretation})}. 
%
After the feature aggregation step, 
each material is represented as a 1-dimensional array with size $m$,
which we interpret as the crystal feature vector containing learned structure-property relationships.
We apply a few fully-connected layers on the crystal feature vector 
and eventually the model will make a multi-class prediction 
from the input point cloud representation.
Other network structures that can deal with 3D equivariance \cite{fuchs2020se} are also viable solutions to our problem, but we find that the performance bottleneck mainly originates from the input features rather than the network, as discussed in \autoref{sec:feature_perturbation}.
To compare the physical knowledge learned by the network, 
we use the same network structure (with different parameters) to learn all target properties.
Details of hyperparameter selection are given in \Autoref{sec:hyperparameter_optimization}.
Model performance in all classification tasks is based on averaging over three independent runs with different random data splits and random seeds.

\section{Results and discussion}

\begin{figure*}[t]
  \centering
  \includegraphics[width=0.98\textwidth]{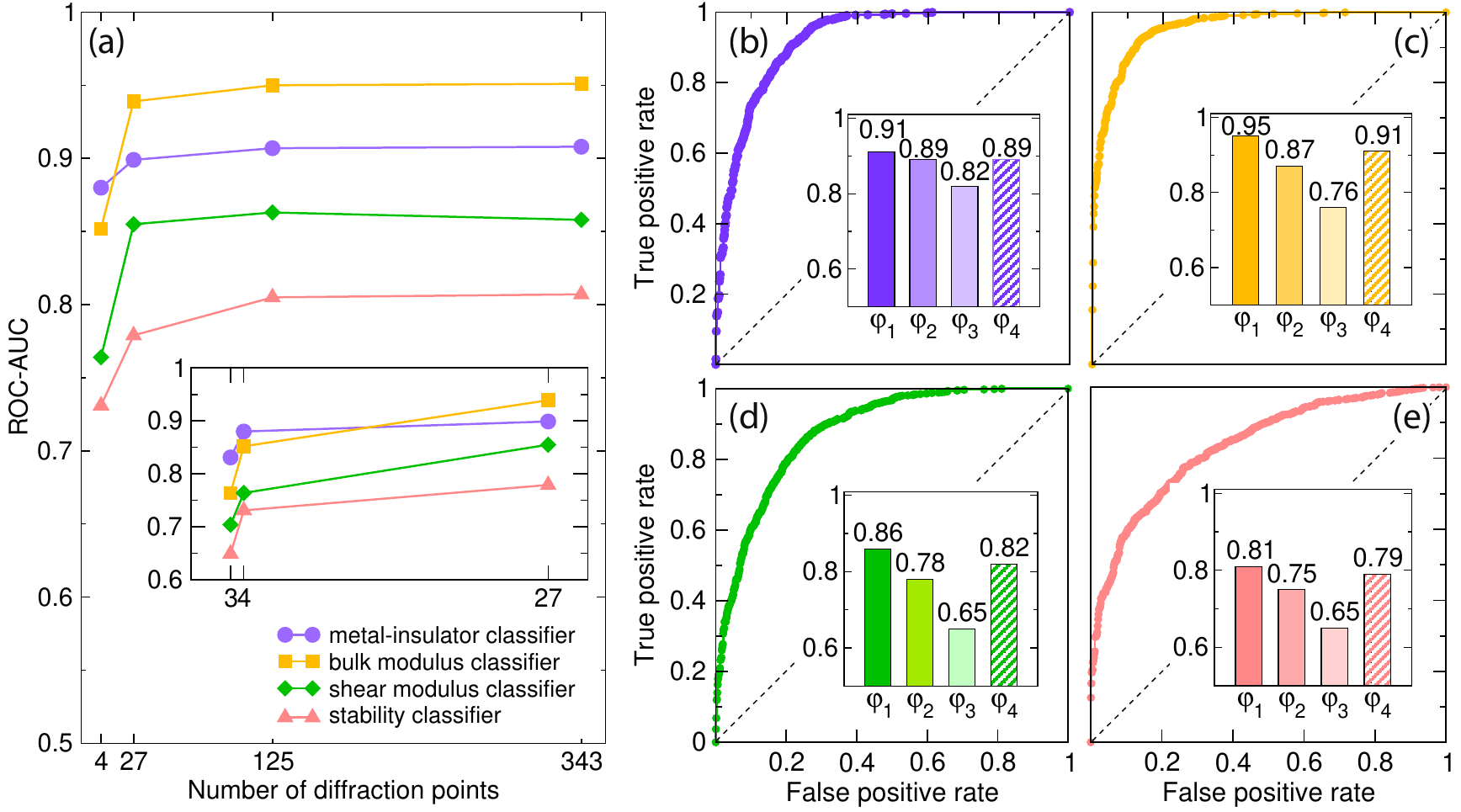}
  \caption{Model performance in multiple classification tasks. (a) ROC-AUC values in four binary classification tasks with a different total number $n$ of diffraction points visible to the model as described in \autoref{sec:materials_representation}.
  ROC curves with $n=343$ for the 
  (b) metal-insulator classification,  
  (c) bulk modulus classification,  
  (d) shear modulus classification, and  
  (e) thermodynamic stability classification. 
  \revision{Insets: Model performance given by ROC-AUC values under various perturbations on input features, with $\varphi_{1}$ -- using the original diffraction patterns, $\varphi_{2}$ -- using randomly scaled intensity, $\varphi_{3}$ -- using systematic absence information, and $\varphi_{4}$ using neutron diffraction patterns, respectively.}
  }
  \label{fig:fig4}
\end{figure*}

\subsection{Crystal System Classification}\label{sec:crystal_system_classification}

We begin our initial assessment of the model learning capability
using a simple computer vision task: crystal-system classification.
The objective is to predict the correct crystal system for a material 
given only the XRD patterns \revision{of its conventional standard cell}.
\revision{This simple task is an important prerequisite before we analyze the structure-property relationship data presented in the next section. If the model is unable to recognize the crystal systems from structural input data, then we may not trust it in making property classifications as they are even more challenging than the current task.}
In order to ensure the quality of crystal data, we consider only materials with cross-reference labels in the
Inorganic Crystal Structure Database (ICSD) database \cite{hellenbrandt2004inorganic}.
We have a total of 48,524 crystal structures from seven crystal systems for this classification task.
Because hexagonal and trigonal cells have identical conventional cell shapes, 
i.e., $a=b\ne c; \alpha=\beta=90^{\circ} \textrm{ and } \gamma=120^{\circ}$, 
we combine these classes together as one, which leads to a total of 6 classes: 
cubic, tetragonal, orthorhombic, hexagonal/trigonal, monoclinic, and triclinic.
Since the crystal systems are uniquely defined by the real space lattice vectors, we only need to provide the model with $n=3$ diffraction points, corresponding to the reciprocal lattice basis vectors (i.e., the given information is ``complete'' for the crystal system classification task). 
We also mask the diffraction intensity information for this task 
by removing the fourth dimension of each point,  
making it invisible to the model.

Our model achieves excellent performance with an accuracy of 0.98 on the test set.
We find that many of the misclassifications are caused by the difference in threshold of ``equivalence.'' 
For instance, the model has difficulty differentiating tetragonal from orthorhombic cells when the ratio of two lattice parameters are approximately unity.
%
%
Furthermore, we tolerate the less-than-perfect accuracy after recognizing 
the network is not fully rotation-invariant for the crystal-system-classification task.
Here, the Euler angles are constrained between $[-\frac{1}{4}\pi, \frac{1}{4}\pi]$ rather than using completely arbitrary rotation angles spanning $2\pi$, 
because the network architecture we use works best with certain spatial orientations of the points.
%
The model has
difficulty in
finding a principal axis and canonicalizing the input when utilizing large rotation angles \cite{qi2017pointnet}.
Nonetheless, it is able to ``visualize'' the shape of the 3D point cloud representation regardless of \revision{small-angle} 3D rotations.
Interestingly, the physical properties considered in the next section are completely immune to random 3D rotations \revision{without any restrictions on Euler angles}, 
which is reasonable as the  properties are scalar quantities \revision{independent of the crystal spatial orientation}.

Next, we ask a \revision{slightly} more challenging question---is it possible to distinguish between materials exhibiting trigonal and hexagonal cells?
\revision{Since this is a binary classification task, we use the area under the receiver operating characteristic curve 
(ROC-AUC, later referred to as AUC) as the performance metric.
The AUC is a value between 0.5 and 1.0, which indicates the model's ability to distinguish a positive case from a negative one.}
We find that given only the three reciprocal lattice basis vectors without diffraction intensity values, 
the model achieves an AUC of 0.87. 
However, once we unmask the diffraction intensity of the three points, 
the AUC value increases to 0.94.
As we further increase the number of diffraction points (with intensity) from $n=3\rightarrow27$, 
the model performance significantly improves.
It distinguishes between the trigonal and hexagonal systems with an $\mathrm{AUC}=0.97$.
The results here primarily show that the \revision{spatial distribution of} diffraction intensity $I$ 
\revision{could further distinguish crystal structures with high similarity,}
which is an advantage of using 3D features over projected 2D patterns \cite{ziletti2018insightful}.
Therefore, we include diffraction intensity information in the following property classification tasks.

\subsection{Property Classification}\label{sec:property_classification}
Given the success of the model to ``visualize'' the shape of the crystal structure, 
we next train our model to classify 
materials properties by learning hidden patterns within the 3D point cloud data based on crystal structure.
The four materials properties we target are metallicity, bulk modulus, shear modulus, and thermodynamic stability.
The binary classification tasks involve: 
($i$) separating compounds without (metals, $E_g\le0$\,eV) from those with a 0\,K gap $E_g>0$\,eV  (insulators) in the electronic structure at the DFT level; ($ii$) 
distinguishing stiff compounds with bulk modulus $B > 85$\,GPa or shear modulus ($G>34$\,GPa, from flexible compounds; and ($iii$) 
 identifying thermodynamically stable materials with $E_H < 20$\,meV\,atom$^{-1}$, respectively.
The fundamental questions that we aim to explore here are twofold:
\begin{itemize}
    \item \textit{Q1}: How much information can we extract from the diffraction patterns for each property classification task?
    \item \textit{Q2}: Which (physical) factors contribute to the classifications, and what is their relative significance?
\end{itemize}

To answer \textit{Q1}, we first examine the impact of the total number of diffraction points ($n$) on model performance for each classification task (\autoref{fig:fig4}a).
For all tasks, we find that as more diffraction points become visible to the model, the performance of the classifier initially improves significantly (from $n=3\rightarrow4\rightarrow27$). The performance then plateaus after 125 points with negligible performance gain using 343 diffraction points.
\autoref{fig:fig4}a also reveals that the electronic band gap and mechanical property classifications in general
have better quality than the thermodynamic stability classification task.
This behavior is reasonable given the importance of composition and chemical identity to material stability \cite{sun2016thermodynamic},
and that the energy above the hull is an ``extrinsic'' property for each phase, whose value is calculated from the energies of other reference phases. 
%

%
Next, we focus on analyzing model performance with smaller $n$ values using the data presented in the inset to  \autoref{fig:fig4}a  where $n=$ 3, 4, and 27.
It is clear that different properties reveal distinct predictive baselines with only three reciprocal basis vectors available ($n=3$).
Specifically, the metal-insulator classifier achieves an AUC of 0.83, a value often considered as an ``effective'' model performance.
Although we typically compare the AUC value of a binary classifier with 0.5 as baseline, here we emphasize in the case of metal-insulator classification, 
one should be cautious in evaluating the model performance, 
since even with very little information available to the model 
as in the $n=3$ case, 
we can still achieve a ``good'' AUC value.
In fact, by analyzing the distribution of metals and insulators across different crystal systems (\autoref{fig:fig6}), we find a high correlation between crystal system and metallicity.
This explains the high classification performance of the structure-based model since we previously showed that it is able to accurately predict the crystal systems from the diffraction patterns.
Moreover, in most materials informatics work, the baseline (i.e., worst-case model performance) is rarely discussed, yet it is quite important for researchers to understand the difficulty of the underlying predictive tasks.
We hope that future materials informatics research could be done in a more physics-first manner, i.e., providing lower and upper bounds to the reported statistical results.

We also note that all models show significant performance gain from $n=3\rightarrow4$, 
where the only difference between these two cases 
is the inclusion of the total electron density information, 
which is the (0\,0\,0) diffraction point at the origin 
with unit phase factor, i.e., $e^0=1$ in \autoref{eq:eq2} and \autoref{eq:eq3}).
In other words, the performance gain from $n=3\rightarrow4$ indicates the quantitative contribution of the total electron density to the classification task (since the origin point does not provide useful spatial information, loosely speaking).
Interestingly, the slopes of performance gain after introducing the origin point for all four classification tasks are similar, despite their different baseline predictive powers.
This finding shows that the total electron density makes a significant contribution to the predictive capability of the model. 

Upon including more diffraction points, i.e., $n\ge27$, the model becomes aware of more symmetry operations in the crystal structure, thus it is reasonable to observe some performance gain as $n$ increases.
Notably, among the curves shown in \autoref{fig:fig4}a, the metal-insulator classifier exhibits the smallest performance gain with increasing $n$, which probably indicates some simpler decision rules underlie the classification compared to the other properties explored. 
Although this may seem counter-intuitive  since determining the electronic band gap is non-trivial, 
the high model performance might be caused by data clustering within the materials database (\autoref{fig:fig6})
owing to the nature of our task being classification rather than regression.
In fact, we did preliminary experiments for regression tasks, e.g., predicting the band gap value using a similar network, 
yet the model showed poor predictive capability, 
possibly due to the simple network architecture being unable to resolve
the complex functions mapping diffraction patterns to materials properties.

\subsection{Perturbing the Input Features}\label{sec:feature_perturbation}
%
To answer \textit{Q2},
we focus on understanding the model performance
for different classification tasks
---what exactly does the model learn from the diffraction patterns?
DNN model interpretability is a known problem owing to the nonlinear activation functions and complex network structures. 
To that end, we choose another route to understand the model performance.
Instead of ``opening the black-box,'' 
we make perturbations to the input features to form new datasets denoted as $\varphi_{i}$, 
and examine the model responses as quantified with the true and false positive rates and ROC-AUC values for each classification task using the same DNN architecture (\autoref{fig:fig4}b-e).
We assign the original diffraction data as $\varphi_{1}$. It contains 
information pertaining to the crystal lattice parameters 
(position of diffraction points), crystal symmetry (spatial distribution of relative diffraction intensity), and electron density (diffraction intensity values).
These are the input features from which we determine the relative 
contributions in the final decision-making of the machine learning model.

To separate the diffraction intensity values from their spatial symmetry,
we generate a random multiplier uniformly sampled within the range (0, 1] for each material during each training iteration, 
and then scale all of its diffraction intensity values with this multiplier before feeding them to the model.
In other words, the random scaling factor differs across different materials as well as each training iteration.
The randomly scaled diffraction patterns correspond to the dataset $\varphi_{2}$, 
which preserve the spatial symmetry (i.e., relative intensities) of the diffraction points, 
but the model would not be able to rely on the absolute values of the intensities, 
which are related to the electron density and atomic numbers (i.e., chemistry).
In addition, we also examined whether the model is learning from systematic absence information in the dataset, 
i.e., $h\,k\,l$ combinations that have zero intensity due to phase cancellation,
to make predictions. 
Dataset $\varphi_{3}$ is obtained by replacing all non-zero diffraction intensity values with unit intensity, $I_{hkl}=1$, while all others remain  $I_{hkl}=0$.
In short, from $\varphi_{1}$ to $\varphi_{3}$, the input features contain a decreasing amount of physical information.
Therefore, by analyzing the classifier performance from various input features, we could infer the relative significance of features to different target properties.

\autoref{fig:fig4}b-e present the model performance with different perturbations to the input diffraction patterns.
We find that the metal-insulator classifier is significantly more robust against random scaling of the intensity values than other classifiers, 
where it is still able to achieve $\mathrm{AUC}=0.89$ with random intensities (see $\varphi_{2}$ in \autoref{fig:fig4}b).
The performance of the bulk modulus and shear modulus classifiers reduce from 0.95  to 0.87, and from 0.86 to 0.78, respectively.
%
%
Notably, we achieve a truly composition-free model after random scaling of the materials diffraction intensity values.
The model completely loses information about atomic number and electron density in this case, 
but it is still aware of which $\mathbf{g}_{hkl}$ points are 
symmetric and their spatial distributions.
Our findings here suggest that the metal-insulator classifier relies mostly on the spatial symmetry of the diffraction patterns, 
while the elasticity-property classifiers depend more on the absolute intensities, which encode the electron density.
All models exhibit inferior performance with only systematic absence information ($\varphi_{3}$ in \autoref{fig:fig4}b-e).
The results here are reasonable, because we further lose some symmetry information as all finite diffraction intensity values become unit intensity.

To further support our previous findings, we perform the same
series of classification tasks using simulated neutron diffraction (ND) patterns.
Unlike X-rays, neutrons interact with the nuclei via the nuclear strong force, whose interaction can be approximated by a short-ranged Fermi pseudopotential.
Since the Fermi pseudopotential is a delta function, whose strength is parameterized by the scattering length $b$, the neutron form factor is $Q$-independent in momentum space, which is the main difference between neutron and X-ray scattering (\autoref{eq:eq3}).
In addition, the neutron scattering lengths are non-monotonic across the periodic table and differ even between isotopes of the same element.
Therefore, the model cannot learn the total electron density in the same way as from the XRD patterns.

The model performance using ND are shown as $\varphi_4$ in \autoref{fig:fig4}b-e.
Interestingly, the model performs slightly worse than using the original XRD features ($\varphi_1$), yet outperforms models with perturbed XRD features ($\varphi_2$ and $\varphi_3$) for all four classification tasks.
By comparing the results from XRD and ND, we are more confident that crystal structure alone plays a significant role in determining the electronic and elastic properties of crystalline materials, while it is less important for thermodynamic stability.

\begin{figure}
  \centering
  \includegraphics[width=0.98\columnwidth]{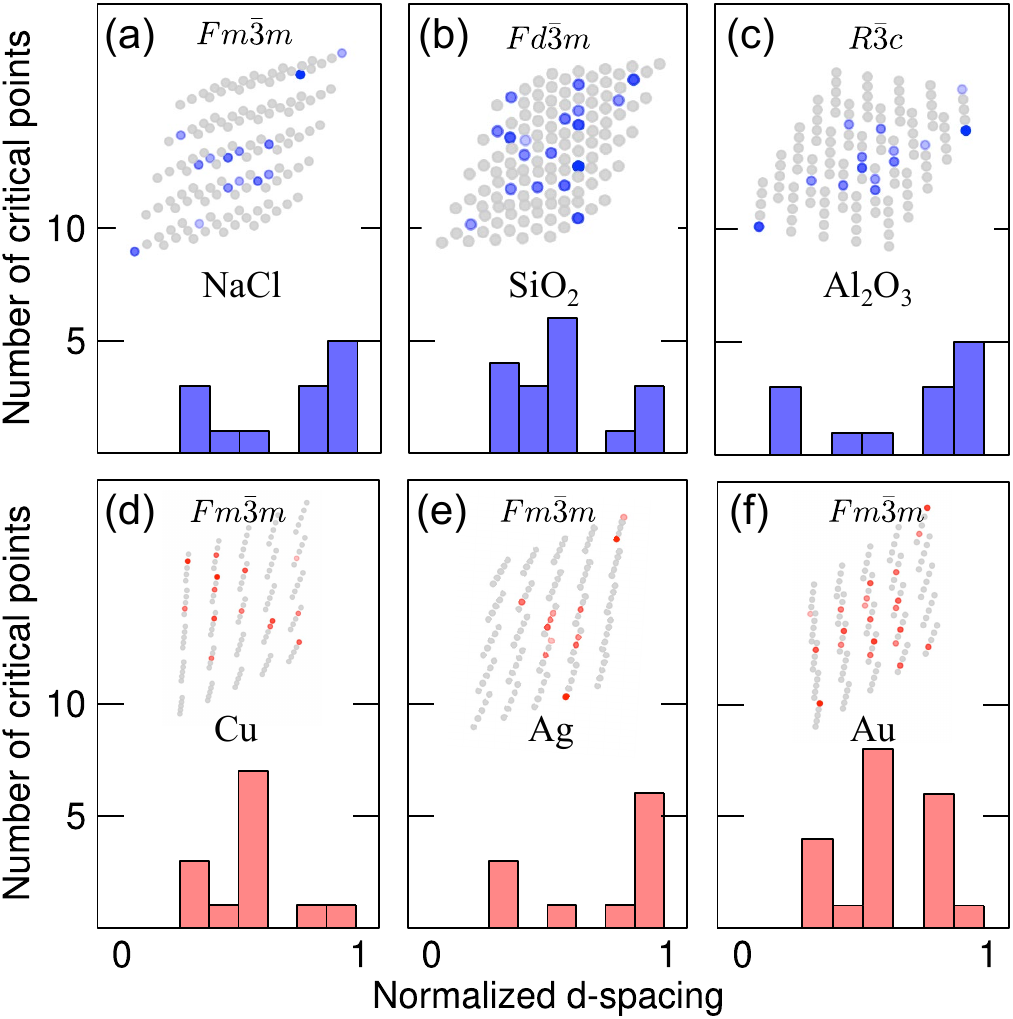}
  \caption{Distribution of critical diffraction points with normalized interplanar $d_{hkl}$ spacings of a few common insulators (NaCl, SiO$_2$, Al$_2$O$_3$) and  metals (Cu, Ag, Au). The $d$-spacings are normalized to facilitate comparison across different materials.
  	The critical points in the limiting sphere (the gray sphere) are those that contribute to the final crystal feature vector after \texttt{max} pooling, and are marked with blue for insulators, and red for metals, respectively.
  	Non-critical points are represented with light gray points.
  	}
  \label{fig:fig5}
\end{figure}

Interestingly, we also built random forest and extreme gradient boosting models based on the diffraction intensities to make metal-insulator classifications for  comparison to the DNN model.
In this case, the diffraction intensities are sorted with respect to their Miller indices (as shown in the bar code in \autoref{fig:fig2}) and they are used as the features for the classifiers.
These ML models are able to achieve quantitatively the same performance (AUC of 0.91) as the DNN model for the given task.
We therefore emphasize that the main objective of this work is not to report a novel neural network with high predictability, but rather to use statistical learning models to understand the materials structure-property relationships.
In the next section, we perform interpretations using the DNN model, 
which is not applicable using the two ML classifiers.
We conclude that both ML and DNN models learn distinct patterns for different target materials properties, 
and they are able to capture physically meaningful features (e.g., spatial symmetry of diffraction patterns) to learn the  materials structure genome and make property predictions.

\subsection{Model Interpretation}\label{sec:model_interpretation}

We now partially open the black box of the DNN model 
to further understand how it classifies metals from insulators.
We plot the distribution of critical points both with normalized interplanar $d_{hkl}$ spacings and in the limiting sphere that contribute to the final crystal feature vector of 6 well-known materials (\autoref{fig:fig5}).
In order to facilitate visualization, we choose a small model which uses $n=125$ diffraction points as input 
and 32-dimensional crystal feature embeddings.
This small model has $\mathrm{AUC}=0.89$, which is acceptable for use in  model interpretation. 
Larger models will have better performance, yet more complicated classification rules.
The model correctly predicts the metallicity of all 6 crystals with high confidence.
The complete list of Miller indices of the critical points are provided in \Autoref{sec:critical_points}.

\revision{We consistently find the inclusion of the origin (0\,0\,0) as a critical diffraction point in all cases, because it contains the total electron density information.}
%
We also find all 6 materials exhibit at least one ``gap'' in the $d$-spacing distribution.
The model relies on lattice planes with large spacings, i.e., small $h$\,$k$\,$l$ points to determine the crystal system,
while also requires information
from the lattice planes
with smaller interplanar distances, 
i.e., higher $h\,k\,l$ indices. 
This observation is reasonable since diffraction points 
with larger Miller indices 
provide more information about 
interplanar interactions, which are 
governed by orbital hybridization 
and attractive and repulsive electrostatic contributions.

We see different critical point distributions for crystals of highly-similar structures with different atomic species (e.g., Cu, Ag, and Au).
A thorough understanding of the model prediction mechanism remains difficult at this time owing to the complicated decision rules underlying the deep neural network.
Interestingly, the model learns the operation of spatial parity. It recognizes inversion symmetry inherent to the XRD patterns (Friedel’s law), 
since it only contains an average of 2 duplicate points with inversion symmetry 
in the final critical point set, e.g., (2\,2\,2) and ($\bar{2}\,\bar{2}\,\bar{2}$).

\begin{table}[t]
\begin{ruledtabular}
\centering
\caption{Model performance for select materials in the perovskite family. `M' and `I' labels indicate metal and insulator, respectively. The score is the probability associated with the predicted class, indicating how confident the model is on that prediction. 
}
\begin{tabular}{lcccc}%
Compound & Space group & True label & Prediction &  Probability\\
\hline
LiNbO$_3$ & $R\bar{3}c$ & I & I & 0.61  \\
LiOsO$_3$ & $R\bar{3}c$ & M & I & 0.51 \\
LaNiO$_3$ & $R\bar{3}c$ & M & I & 0.56  \\
LaCoO$_3$ & $R\bar{3}c$ & M & M & 0.51 \\
LiNbO$_3$ & $R3c$ & I & I & 0.64  \\
LiOsO$_3$ & $R3c$ & M & M & 0.54  \\
LiTaO$_3$ & $R3c$ & I & I & 0.59   \\
\hline
NdNiO$_3$ &  $Pnma$ & M & I & 0.58 \\
YNiO$_3$ & $Pnma$ & M & I & 0.59 \\
CaFeO$_3$ & $Pnma$ & M & I & 0.84 \\
SrRuO$_3$ & $Pnma$ & M & I & 0.89  \\
CaTiO$_3$ & $Pnma$ & I & I & 0.91 \\
\hline
NdNiO$_3$ &  $P2_{1}/c$ & I & M & 0.51  \\
YNiO$_3$ & $P2_{1}/c$ & I & I & 0.55  \\
CaFeO$_3$ & $P2_{1}/c$ & I & I & 0.80  \\
\hline
SrFeO$_3$ & $Pm\bar{3}m$ & M & M & 0.80  \\
SrTiO$_3$ & $Pm\bar{3}m$ & I & M & 0.73
\end{tabular}\label{tab:perovskite}
\end{ruledtabular}
\end{table}

\subsection{Limitations of Structure-based Learning}

Since we do not explicitly have elemental composition information in the XRD patterns, we expect the model to have difficulty making predictions on materials from the same family, i.e., with similar crystal structures yet different compositions and various properties.
To that end, we examine the performance of our structure-based model on the $AB$O$_3$ perovskite family (\autoref{tab:perovskite}).
All compounds listed  here were removed from the training and validation dataset for this classification task. 

Overall the model performs poorly in classifying metals from insulators in the perovskite family.
We find the model tends to predict 
all orthorhombic $Pnma$ 
compounds to be insulators.
However, \Autoref{fig:database_stats} shows that metals and insulators are quite balanced for the orthorhomic crystal system.
It is therefore not clear why our model exhibits such behavior.
The model in general exhibits low confidence scores in predicting most of the perovskite materials, which is reasonable since minor structural distortions in these materials could drive metal-to-insulator transitions \cite{RevModPhys.70.1039}, while the change in diffraction patterns might be indistinguishable to the model. 
The model performance in the perovskite family is reasonable since undoubtedly chemistry and interactions among different microscopic electronic, spin, and orbital degrees-of-freedom play a significant role in determining materials properties.
Although the perovskite famility poses a challenge to the structure-based model, 
the poor performance is expected since we designed this task to reveal the limitations of only using structural information to predict materials properties.
The aforementioned model performance across many structure types 
still uncovers that metals and insulators exhibit distinct XRD patterns, and our model is able to capture those differences effectively.

\section{Conclusions and outlook}

In conclusion, we use DNN models to show the intimate correlation between crystal structure and multiple materials properties. 
We learn from both XRD and ND patterns that crystal symmetry plays a significant role in determining electronic band gaps, 
while electron density contributes more to elastic properties.
Stability, however, is more composition-dependent and therefore our model exhibits poorest performance in predicting this thermodynamic response.
These findings impart a better understanding of the role of crystal structures in functional materials properties.

Moreover, if we have the exact Fourier series expansion of the periodic electron density function in real space, 
it would be theoretically possible to construct a sophisticated enough DNN model to learn the functionals that map ground state electron density to materials properties.
However, this would require us to obtain orders of magnitude more 
number of points instead of only a few hundred, which is currently impractical.
Based on our current understanding, 
the neural network is most likely not learning the functional mapping, 
but mainly making predictions based on spatial symmetry and electron density information hidden in the diffraction patterns.
In other words, it is performing complex pattern recognition rather than learning the underlying functional relationship and mathematical structure of materials.
This fact may be a result of performing classification tasks rather than regression modeling.
We suspect that learning the density functional mapping using a regression DNN model is possible, but requires a large neural network of unknown architecture.

Lastly, our work here not only reveals some interesting correlation between crystal structure and materials properties, 
but also demonstrates the capability of statistical learning models 
beyond making accurate property predictions/classifications.
We argue that they are also valuable in advancing our materials-physics understanding through statistical analysis.
With a properly designed learning model and physically meaningful feature perturbations, it is possible to utilize machine learning models in a similar manner to first principles methods to study system response upon perturbations.
While first principles models can handle an individual system well, machine learning could reveal the hidden patterns buried underneath a large amount of experimental/simulation data.
This makes statistical learning a complementary method to theoretical modeling and physics-based simulations towards a better understanding of materials physics.

\section*{Data Availability}

The data that support the findings of this study are also available from the corresponding author upon reasonable request. \revision{The same results should also be reproducible using up-to-date Materials Project database.} Our model is open-sourced on \href{https://github.com/raymond931118/deepKNet}{GitHub} \cite{code-link} with automatic data acquisition/processing pipeline and model training instructions.

\begin{acknowledgments}

Y.W.\ and X.Z.\ developed the theoretical framework and implemented the model. F.X.\ helped design the neural network and automated the learning process. J.M.R. conceived and administered the project. E.A.O.\ contributed to the network architecture optimization, R.S.\ and S.D.W.\ guided the physical interpretation of the statistical learning model. We thank Dr.\ Danilo Puggioni at Northwestern University for helpful discussions. 
\revision{The authors also thank the anonymous reviewers for their helpful
and constructive comments.}

This work was supported in part by the National Science Foundation (NSF) under award number DMR-1729303 (Y.W.\ and J.M.R.) and DMR-1729489 (R.S.\ and S.D.W.).
The information, data, or work presented herein was also funded in part by the Advanced Research Projects Agency-Energy (ARPA-E), U.S.\ Department of Energy, under Award Number DE-AR0001209 (E.A.O.).
The views and opinions of authors expressed herein do not necessarily state or reflect those of the United States Government or any agency thereof.

\end{acknowledgments}

\appendix
\renewcommand{\thefigure}{A\arabic{figure}}
\renewcommand{\thetable}{A\arabic{table}}
\setcounter{figure}{0}
\setcounter{table}{0}

\section{X-ray and Neutron Diffraction Simulations} \label{sec:derivations}

We use a modified version of the X-ray diffraction (XRD) simulator as implemented in the open-source software \texttt{Pymatgen} \cite{ong2013python} to generate the input features for our model. Specifically, we consider all diffraction points within the limiting sphere of radius $4\pi/\lambda$, where $\lambda=1.5418$\,\AA\ is the X-ray wavelength (Cu\,K$_\alpha$ in this case). The atomic form factors $f(s)$ are calculated using tabulated data to simulate the Fourier-transformed real-space atomic electron density function $\rho(\mathbf{r})$:
\begin{equation}
    f(s) = Z - 41.78214 \cdot s^2 \cdot \sum_{i=1}^{n} a_i e^{-b_i s^2}\,, \label{eq:appendix_eq1}
\end{equation}
where $Z$ is the atomic number, 
$s = \frac{\sin\theta}{\lambda}$ and $\sin\theta = \frac{\lambda}{2d_{hkl}}$ (Bragg condition). The $a_i$ and $b_i$ coefficients are $n$ fitting parameters for each element provided by \texttt{Pymatgen}. We then calculate
\begin{eqnarray}
    F_{hkl} &=& \sum_{j=1}^N f_j e^{2\pi i \mathbf{g}_{hkl}\cdot \mathbf{r_j}} \label{eq:appendix_eq2} \\
    I_{hkl} &=& \frac{1}{V_{cell}^2} F^{*}_{hkl} F_{hkl} \label{eq:appendix_eq3}
\end{eqnarray}
where $j$ runs over all atoms within the unit cell and $V_{cell}$ is the unit cell volume (either primitive or conventional). Lorentz polarization and Debye-Waller factor are not considered in our simulation. Owing to the large values of the diffraction intensity, we take the natural logarithm of each $I_{hkl}$, i.e. $\tilde{I}_{hkl} = \ln(I_{hkl}+10^{-6})$ and normalize the intensity values to range (0, 1].

For neutron scattering, \autoref{eq:appendix_eq1} becomes a constant for each element (more specifically, for each isotope), which is independent of the momentum space position vector. The tabulated neutron scattering lengths are obtained from the \texttt{Pymatgen} package \cite{ong2013python}. We then calculate the neutron diffraction (ND) patterns as before using \autoref{eq:appendix_eq2} and \autoref{eq:appendix_eq3}, after obtaining the neutron scattering lengths.
\revision{The adapted simulation code can be found at the project  \href{https://github.com/raymond931118/deepKNet}{GitHub} page \cite{code-link}.}

\section{Hyperparameter Optimization} \label{sec:hyperparameter_optimization}

The hyperparameters we considered for the DNN model are tabulated in
\autoref{tab:hyperparameters_DL}. Note that the number of diffraction points $n$ is considered as a variable instead of a hyperparameter of the model. We use a greedy approach to optimize each of these hyperparameters and take the average results from three randomly and independently generated training, validation, and test datasets.
The reported data was generated using MLPs with dimension $[64, 128, 512, 1024]$, the \texttt{max} pooling function, and then MLPs with hidden dimension of $[1024, 512, 256, 128, k]$, where $k$ is the number of output class. However, we find that the model performance on all classification tasks to be less dependent on the hyperparameters than the input perturbations. In other words, the same neural network architecture could capture most of the materials information, and increasing the number of model parameters does not significantly improve the model performance.

The hyperparameters for the machine learning models are tabulated in \autoref{tab:hyperparameters_ML}. We obtained almost quantitatively identical results using random forest and extreme gradient boosting algorithms for the metal-insulator classification task.

\begingroup
\squeezetable
\begin{table}[h]
\begin{ruledtabular}
\centering
\caption{Hyperparameters explored in construction of the DNN model.
}
\begin{tabular}{ll}%
Hyperparameter & Values \\
\hline
number of first MLP layers & 3, 4, 5 \\
MLP hidden dimension & 64, 128, 256, 512, 1024 \\
dimension of hidden crystal feature vector & 256, 512, 1024 \\
number of fully-connected layers & 3, 4, 5 \\
size of fully-connected layers & 128, 256, 512, 1024 \\
pooling & max, self-attention \\
number of self-attention layers & 0, 2, 4 \\
optimizer & SGD, Adam \\
initial learning rate & 0.01, 0.001 \\
dropout & 0, 0.2, 0.4 \\
\end{tabular}\label{tab:hyperparameters_DL}
\end{ruledtabular}
\end{table}
\endgroup

\begin{table}[t]
\begin{ruledtabular}
\centering
\caption{Hyperparameters explored in the random forest and extreme gradient boosting classifiers using the \texttt{scikit-learn} package.
}
\begin{tabular}{ll}%
Hyperparameters & Values \\
\hline
number of estimators & 50, 100, 150 \\
criterion & gini, entropy \\
maximum depth & 6, 12, None \\
maximum number of features & None, sqrt, log2 \\
\end{tabular}\label{tab:hyperparameters_ML}
\end{ruledtabular}
\end{table}

\section{Standard Deviation of Property Learning Tasks} \label{sec:standard_deviations}
\autoref{tab:stdevs} shows the standard deviations for each property classification task from 3 randomly and independently generated training-validation-test datasets. 
\begin{table}[t]
\begin{ruledtabular}
\centering
\caption{Standard deviations from 3 randomly and independently generated training-validation-test datasets for different property classification tasks.}
\begin{tabular}{lllll}
Classification Task & $\varphi_1$ & $\varphi_2$ & $\varphi_3$ & $\varphi_4$  \\
 \hline
metallicity & 0.002  &  0.001 & 0.005 & 0.005  \\
 \hline
bulk modulus & 0.002  & 0.004 & 0.011 & 0.001 \\
 \hline
shear modulus & 0.011 & 0.017 & 0.014 & 0.003 \\
 \hline
stability & 0.003 & 0.006 & 0.004 & 0.004 \\

\end{tabular}\label{tab:stdevs}
\end{ruledtabular}
\end{table}

\section{Database Statistics}
\begin{figure}[h]
  \centering
  \includegraphics[width=0.99\columnwidth]{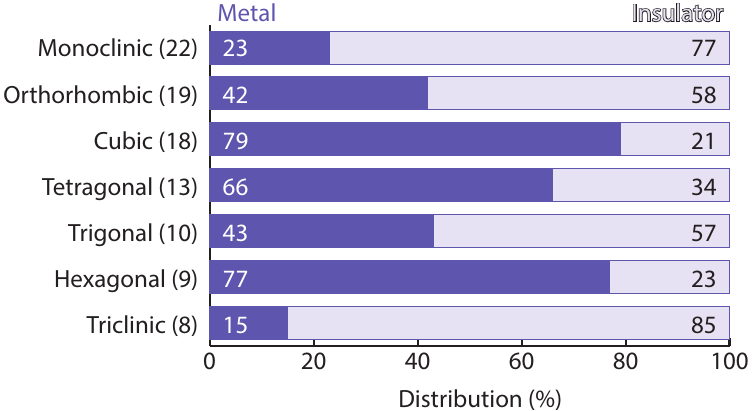}
  \caption{The distribution of metals and insulators across various crystal systems within our dataset. The total number of crystal structures is 74,108.}
  \label{fig:fig6}\label{fig:database_stats}
\end{figure}

\section{Critical Points}\label{sec:critical_points}
We use 125 diffraction points (Miller indices within \{$\bar{2}$, $\bar{1}$, 0, 1, 2\}) for all materials and 32-dimensional crystal feature embedding for the model interpretation task. The critical points of each material shown in \autoref{fig:fig5} are enumerated below. Since the same point may contribute to more than one hidden crystal feature, the number of unique critical points could be less than the embedding dimension. \\

\noindent Critical points of NaCl:
\begin{enumerate}
	\item[] (0 0 0) \quad (0 0 1) \quad (0 0 $\bar{1}$) \quad (0 $\bar{1}$ 0) \quad (1 0 0)
	\item[] (0 $\bar{1}$ $\bar{1}$) \quad (1 0 1) \quad ($\bar{1}$ $\bar{1}$ 0) \quad (1 1 2) \quad ($\bar{1}$ $\bar{1}$ $\bar{1}$)
	\item[] ($\bar{1}$ $\bar{2}$ 1) \quad (2 2 $\bar{2}$) \quad (2 $\bar{2}$ 2) \quad ($\bar{2}$ 2 $\bar{2}$)
\end{enumerate}

\noindent Critical points of Al$_2$O$_3$:
\begin{enumerate}
	\item[] (0 0 0) \quad (0 0 $\bar{1}$) \quad (0 1 0) \quad (0$\bar{1}$ 0) \quad (1 0 0)
	\item[] (1 0 1) \quad (1 1 0) \quad ($\bar{1}$ 1 0) \quad ($\bar{1}$ $\bar{1}$ 0) \quad (2 0 2)
	\item[] (1 1 1) \quad (2 $\bar{2}$ 2) \quad (2 $\bar{2}$ $\bar{2}$) \quad ($\bar{2}$ 2 $\bar{2}$)
\end{enumerate}

\noindent Critical points of SiO$_2$:
\begin{enumerate}
	\item[] (0 0 0) \quad (0 0 $\bar{1}$) \quad (1 0 0) \quad ($\bar{1}$ 0 0) \quad ($\bar{2}$ 0 0)
	\item[] (0 2 $\bar{2}$) \quad (0 $\bar{2}$ 2) \quad (1 0 1) \quad (1 0 $\bar{1}$) \quad ($\bar{1}$ 1 0)
	\item[] (2 0 2) \quad (2 0 $\bar{2}$) \quad ($\bar{2}$ 0 2) \quad ($\bar{1}$ $\bar{1}$ 1) \quad ($\bar{1}$ $\bar{1}$ $\bar{2}$)
	\item[] ($\bar{1}$ 2 1) \quad ($\bar{1}$ $\bar{2}$ $\bar{1}$) \quad ($\bar{2}$ $\bar{1}$ $\bar{2}$)
\end{enumerate}

\noindent Critical points of Ag:
\begin{enumerate}
	\item[] (0 0 0) \quad (0 0 $\bar{1}$) \quad (0 $\bar{1}$ 0) \quad (1 0 0) \quad ($\bar{1}$ 0 0)
	\item[] (0 $\bar{1}$ $\bar{1}$) \quad (1 1 1) \quad ($\bar{1}$ $\bar{1}$ $\bar{1}$) \quad (2 2 $\bar{2}$) \quad (2 $\bar{2}$ $\bar{1}$)
	\item[] ($\bar{2}$ $\bar{1}$ 0) \quad ($\bar{2}$ 2 $\bar{2}$)
\end{enumerate}

\noindent Critical points of Cu:
\begin{enumerate}
	\item[] (0 0 0) \quad (0 $\bar{1}$ 1) \quad (1 0 1) \quad (1 $\bar{1}$ 0) \quad ($\bar{1}$ 0 1)
	\item[] (2 0 1) \quad (1 1 2) \quad (1 1 $\bar{2}$) \quad (1 $\bar{2}$ 1) \quad ($\bar{1}$ $\bar{1}$ $\bar{1}$)
	\item[] ($\bar{1}$ $\bar{1}$ $\bar{2}$) \quad ($\bar{1}$ $\bar{2}$ $\bar{1}$) \quad (2 $\bar{1}$ 2) \quad ($\bar{2}$ $\bar{2}$ $\bar{2}$)
\end{enumerate}

\noindent Critical points of Au:
\begin{enumerate}
	\item[] (0 0 0) \quad (0 0 $\bar{1}$) \quad (1 0 1) \quad ($\bar{1}$ 0 1) \quad ($\bar{1}$ 0 $\bar{1}$)
	\item[] ($\bar{1}$ $\bar{1}$ 0) \quad (0 $\bar{1}$ $\bar{1}$) \quad (0 1 1) \quad (0 1 $\bar{1}$) \quad (1 0 2)
	\item[] (1 1 0) \quad (1 $\bar{1}$ 0) \quad ($\bar{1}$ 1 0) \quad ($\bar{2}$ 1 0) \quad ($\bar{1}$ 2 $\bar{2}$)
	\item[] ($\bar{1}$ $\bar{2}$ $\bar{1}$) \quad ($\bar{1}$ $\bar{2}$ $\bar{2}$) \quad (2 2 $\bar{2}$) \quad (2 $\bar{2}$ 2) \quad ($\bar{2}$ $\bar{1}$ $\bar{1}$)
	\item[] ($\bar{2}$ 2 2)
\end{enumerate}

\bibliography{reference}

\end{document}